\documentclass[aps,prb,floatfix,superscriptaddress,twocolumn,nofitinbib,hyperref=pdftex,citeautoscript]{revtex4}
\usepackage{epsfig}

\usepackage{amsmath,bm}
\usepackage{amssymb}
\usepackage{braket}
\usepackage{amsfonts}
\usepackage{dsfont}
\usepackage{comment,color}
\usepackage{lipsum}
\usepackage{hyperref}

\hypersetup{colorlinks=true,linkcolor=blue,anchorcolor=blue,citecolor=blue,filecolor=blue,urlcolor=blue,bookmarksnumbered=true,pdfview=FitB}

\usepackage[dvipsnames]{xcolor}

\begin{document}
 \title{Floquet multiple exceptional points with higher-order skin effect}

\author{Gaurab Kumar Dash}
\affiliation{Department of Physics, Indian Institute of Technology Delhi, Hauz Khas, New Delhi, India 110016}
\author{Subhendu Kumar Patra}
\affiliation{Department of Physics, Indian Institute of Technology Hyderabad, Kandi, Sangareddy, Telengana, India 502285}
\author{Diptiman Sen}
\affiliation{Centre for High Energy Physics, Indian Institute of Science, Bengaluru, Karnataka 560012, India}
\author{Manisha Thakurathi}
\affiliation{Department of Physics, Indian Institute of Technology Hyderabad, Kandi, Sangareddy, Telengana, India 502285}

\date{\today}
\begin{abstract}
    We investigate the rich non-equilibrium physics arising in periodically driven open quantum systems, specifically those realized within microcavity resonators, whose dynamics are governed by a non-Hermitian Hamiltonian hosting Floquet Exceptional Points (FEPs). By introducing a periodically quenched driving protocol, we analytically derive the Floquet effective Hamiltonian and determine the locations of multiple FEPs harbored within the Floquet bulk bands. We demonstrate that the pair-production and annihilation of these FEPs can be precisely controlled by fine-tuning the system parameters, and zero and $\pi$ FEPs are topologically characterized by robust integer quantized winding numbers. To probe these singularities, we introduce a bi-orthogonal Floquet fidelity susceptibility, whose value exhibits large non-zero peaks at the momentum points hosting FEPs in the Brillouin zone. Furthermore, the momentum-summed susceptibility displays a sharp divergence when the number of FEPs change with respect to the time period of the drive. Our findings also reveal the emergence of Floquet edge states around zero energy and Dirac-like dispersion around $\pi$. Moreover, our model reveals a higher-order skin effect, where the periodically driven Hamiltonian hosts skin modes localized at both edges and corners. These insights offer novel avenues for the Floquet engineering of topological singularities in driven dissipative systems, with significant potential for manipulating light and matter at the microscale.
\end{abstract}
\maketitle
\section{Introduction}
Periodic driving of a physical system offers a powerful means of dynamically modifying its properties, enabling the realization of novel phenomena and phases of matter that lack static counterparts [\onlinecite{floquet-markov, oka, PhysRevB.89.235434, chaudhury2009quantum,PhysRevB.88.155133,PhysRevResearch.2.013292}]. This Floquet engineering approach, particularly under high-frequency driving, can induce topologically non-trivial band structures by dynamically breaking relevant symmetries. In these driven systems, quasi-energies ($\epsilon$) are defined modulo $\hbar \omega$ (with $\omega=2 \pi/T$ and $T$ being the time-period of the driving), allowing Floquet bands to connect across the boundaries of the Floquet Brillouin zone (FBZ), $\epsilon \in [-\hbar \omega/2, \hbar \omega/2)$ and giving rise to Floquet diabolic points [\onlinecite{lindner2011floquet,ho2012quantized, cayssol2013floquet, bukov2015universal, PhysRevResearch.1.032013, holthaus2015floquet, eckardt2017colloquium, weinberg2017adiabatic, oka2019floquet, rudner2020band,PhysRevB.95.155407}].

Recent years have witnessed significant advances in the study of non-Hermitian (NH) topological phases, with phenomena such as exceptional points (EPs) and the non-Hermitian skin effect (NHSE) attracting considerable attention [\onlinecite{EP8,gkd1, gkd2, gkd3, gkd4, EP1, EP2, EP3, EP4, EP5, EP6, EP7, NHSE1, NHSE2, NHSE3, NHSE4, NHSE5, NHSE6, NHSE7, se2, se3, se5}]. An EP represents a unique NH band degeneracy where both the eigenvalues and the corresponding eigenvectors coalesce, such that the algebraic and geometric multiplicities of the Hamiltonian become unequal. This has led to a wide variety of semimetallic NH phases, where nodal degeneracies generalize into knots, lines, loops, and surfaces formed by EPs [\onlinecite{k1, k2, k3, k4, k5, k6}]. Experimentally realizable NH systems, often exhibiting parity-time ($\mathcal{P}\mathcal{T}$) symmetry due to balanced loss and gain, have been successfully classified using novel schemes that extend beyond their Hermitian counterparts [\onlinecite{pt1, pt2, pt3, pt4}]. In NH systems, the concept of fidelity susceptibility serves as a powerful tool to characterize critical phenomena and, crucially, to detect EPs [\onlinecite{nfs1, nfs2, nfs3, nfs4, nfs5, nfs6, nfs7}]. Unlike Hermitian systems, where fidelity susceptibility typically involves the overlap of a state with itself under a small parameter change [\onlinecite{fs1, fs2, fs3, fs4, fs5, PhysRevB.86.245424}], NH systems require the use of biorthogonal fidelity susceptibility. This generalization accounts for the non-orthogonality of eigenvectors in NH Hamiltonians by utilizing both left and right eigenvectors. In an EP, the denominator in the expression of biorthogonal fidelity susceptibility approaches zero, leading to a characteristic divergence in its value. This divergence often manifests itself as a sharp peak or singularity in the parameter space, allowing for the precise pinpointing of EP locations through the monitoring of fidelity susceptibility as system parameters are tuned.

Microcavity resonators provide an ideal experimental platform for realizing these NH phenomena [\onlinecite{mc1, mc2, mc3}]. In these systems, light circulates in degenerate clockwise and anticlockwise modes. By introducing precisely controlled on-site loss in one mode and on-site gain in the other, the system's effective Hamiltonian becomes NH, allowing for the formation of EPs when parameters are tuned to induce eigenvalue coalescence [\onlinecite{mc5, mc6, mc7, mc8, mc9, mc10, mc11, mc12, mc13}]. This concept extends to the Floquet domain: applying a periodic drive to such an NH microcavity leads to a Floquet Hamiltonian, whose interplay with non-Hermiticity can result in the emergence of FEPs in the quasi-energy spectrum, controllable by tuning both gain/loss and driving parameters. Furthermore, the NHSE highlights the extreme sensitivity of NH systems to boundary conditions, causing bulk states to accumulate at the system's edges, fundamentally altering the traditional bulk-boundary correspondence. More recently, the discovery of higher-order skin effects has revealed new types of boundary modes localized at corners and other boundaries of higher codimensions, different from conventional topological insulators [\onlinecite{NHSE1, NHSE2, NHSE3, NHSE4, NHSE5, NHSE6, NHSE7}].

Therefore, the intricate interplay between periodically driven topological phases and NH phases of matter, encompassing FEPs and the NHSE, has become a central focus in condensed matter physics. Despite the current surge in research on higher-order skin effects and periodic driving, their combined effects along with their detection using fidelity susceptibility remain largely unexplored. In this article, we investigate this interplay by introducing a periodically quenched driving protocol in a two-dimensional, two-band NH system realized in an array of microcavity resonators. We show that the static counterpart of the model showcases EPs with integer-quantized winding numbers, and their defining property in NH systems is confirmed by the sharp divergence of the absolute value of biorthogonal fidelity susceptibility (BFS). For the driven system, the Floquet effective Hamiltonian harbors multiple FEPs in the FBZ, whose pair-production and annihilation can be controlled by fine-tuning the system parameters. These FEPs are topologically characterized by robust integer quantized winding numbers around zero and $\frac{\pi}{2T}$ quasi-energy. We introduce bi-orthogonal Floquet fidelity susceptibility (BFFS) to characterize the FEPs in the system, demonstrating that momentum points hosting FEPs exhibit non-zero peaks. Remarkably, the momentum-summed absolute value of this susceptibility displays a sharp divergence when the number of FEPs changes as a function of the time period, confirming their generation and annihilation. The model, remarkably, showcases the emergence of $\pi$ surface Dirac cones in the open boundary spectra. Along with this, the complex eigenspectrum of the effective Hamiltonian exhibits Floquet point gaps, lacking static counterparts under periodic boundary conditions (PBC). These are accompanied by higher-order skin modes localized at the corners and along the edges under open boundary conditions in both axes, thereby revealing a higher-order skin effect in periodically driven NH systems.

Our investigation is presented in a structured manner. In Section II, we first introduce a static model designed to harbor EPs within the Brillouin zone (BZ), notably featuring divergent BFS. This is followed by Section III, which examines the established emergence of FEPs in perodically driven NH systems. Section IV then proceeds to unveil a periodically quenched model, showcasing the emergence of multiple FEPs with divergent BFFS and their association with higher-order NHSE. And finally we conclude in Section V. For further illustration, the appendix details the presence of multiple EPs within the context of the NH Qi-Wu-Zhang (QWZ) model.

\section{Two band model with multiple EPs}
We introduce a two-band model that exhibits EPs in the BZ which can be described by the following model Hamiltonian [\onlinecite{gkd2,gkd3}]
\begin{figure}
    \centering
    \includegraphics[width=1\linewidth]{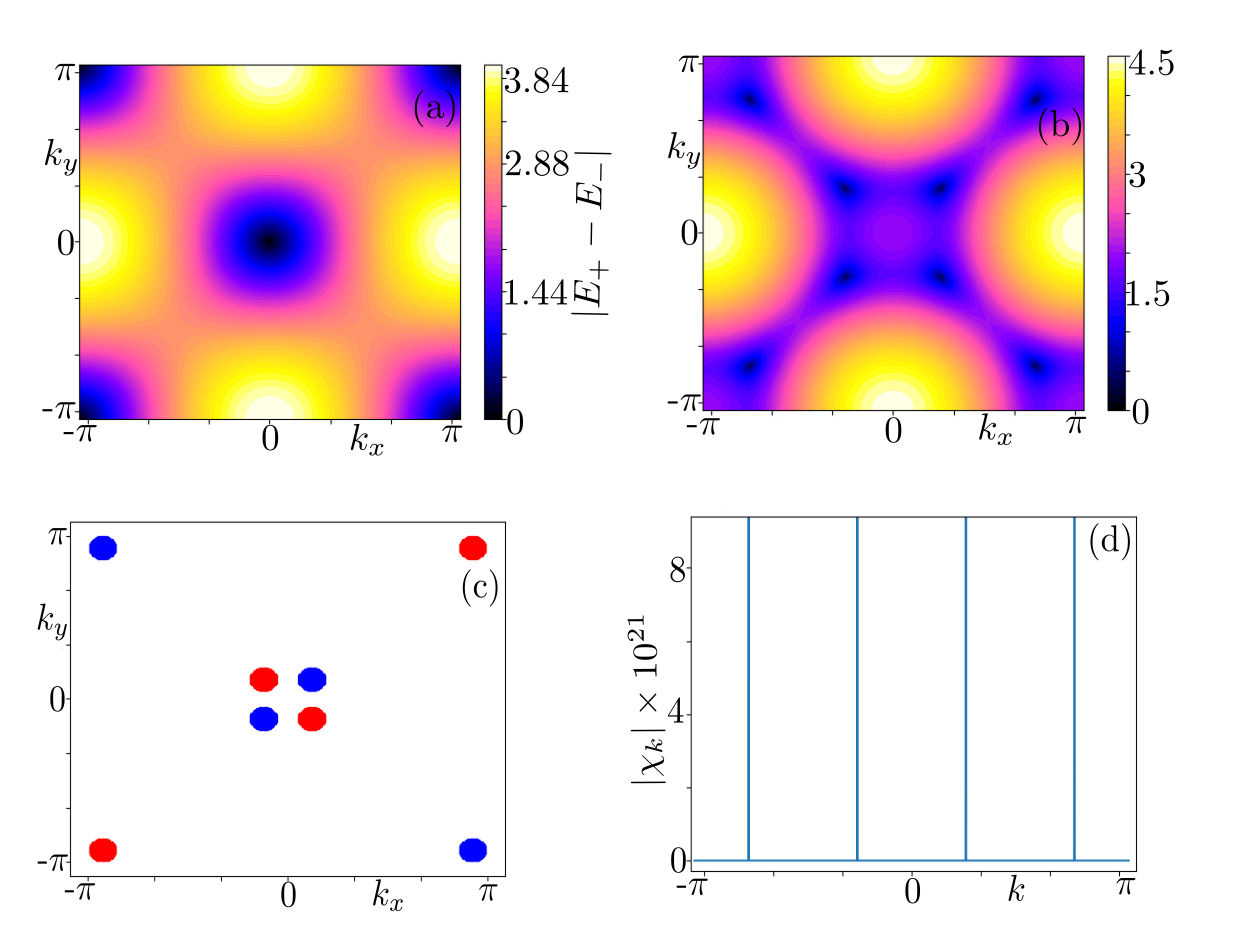}
    \caption{(a) showcases the contour plot of absolute part of the energy difference $\lvert E_{+} - E_{-}\rvert$ of the Hermitian model ($\xi=0 $ in Eq. (\ref{eqn1})) across the Brillioun zone illustrating the presence of five Dirac points under periodic boundary conditions. (b) illustrates the emergence of EPs in the NH regime. (c) displays the integer quantized winding number $W$ of EPs color coded as red($+1$) and blue($-1$). (d) presents the biorthogonal fidelity susceptibility as a function of momentum $k$ (where $k=k_x = k_y$) highlighting its divergence at the EP. The model parameters are fixed at $\xi/\lambda= 1$ and $\Delta/\lambda=1$ for all presented figures specifically for (b)-(d).}
    \label{fig1}
\end{figure}

\begin{equation}
    \mathcal{H} = \lambda(\sin{k_x}\sigma_y - \sin{k_y}\sigma_x) +\left[ \Delta(\cos{k_x}-\cos{k_y}) + i\xi\right]\sigma_z .
\label{eqn1}
\end{equation}
The Hamiltonian, in the Hermitian regime ($\xi=0$), is characterized by explicit breaking of both the time-reversal symmetry ($\mathcal{T}=-i\sigma_y\mathcal{K}$, where $\mathcal{K}$ denotes complex conjugation) and rotational symmetry $\mathcal{C}_{4z}=e^{i\frac{\pi}{4}\sigma_z}$ while preserving the combined symmetry $\mathcal{C}_{4z}\mathcal{T}$ [\onlinecite{ams}]. However, the introduced non-hermiticity breaks $\mathcal{C}_{4z}\mathcal{T}$. The NH model respects the mirror symmetries associated with transposition along the $x$ and $y$ directions, which is given by
\begin{equation}
   \mathcal{M}_x\mathcal{H}^{T}(k_x,k_y)\mathcal{M}^{-1}_{x}= \mathcal{H}(-k_x,k_y),
\end{equation}
and
\begin{equation}
   \mathcal{M}_y\mathcal{H}^{T}(k_x,k_y)\mathcal{M}_y^{-1}= \mathcal{H}(k_x,-k_y),
\end{equation}

where $\mathcal{M}_x=\sigma_0$ and $\mathcal{M}_y=\sigma_z$. The combination of the mirror symmetries imply the existence of the spatio-inversion symmetry $\mathcal{P}(=\mathcal{M}_x\mathcal{M}_y)$
\begin{equation}
    \mathcal{P}\mathcal{H}(k_x,k_y)\mathcal{P}^{-1} = \mathcal{H}(-k_x,-k_y).
\end{equation}

This Hamiltonian is amenable to experimental realization within a microcavity resonator system. In this context, the $\sigma$ matrices describe the coupling between the clockwise and anticlockwise modes within the coupled cavity resonator. Furthermore, the non-Hermiticity can be generated by introducing on-site loss in the clockwise modes and on-site gain in the anticlockwise modes, respectively. Such precisely controlled modifications to potentials are experimentally attainable in coupled microcavity resonators via asymmetric scattering between their respective clockwise and anticlockwise modes [\onlinecite{as1, as2}].

Next, we calculate the complex band dispersion written as

\begin{equation}
  E_{\pm}(\mathbf{k}) = \pm\sqrt{\alpha(\mathbf{k})} = \pm\sqrt{\alpha_R(\mathbf{k}) + i\alpha_I(\mathbf{k})},
\label{eqn5}
\end{equation}
where the real and imaginary components of $\alpha(\mathbf{k})$ are defined as: $\alpha_R(\mathbf{k}) = \lambda^2\sum_{i=x,y}\sin^2{k_i} + \Delta^{2}(\cos{k_x}-\cos{k_y})^2 - \xi^2$
and $\alpha_I(\mathbf{k}) = 2\Delta\xi(\cos{k_x}-\cos{k_y})$.
Here, $\mathbf{k} = (k_x,k_y)$ denotes the crystal momentum within the first BZ. In the Hermitian limit ($\xi=0$), the model hosts Dirac points at the $\mathcal{C}_{4z}\mathcal{T}$ symmetric momentum points $(0,0)$ and $(\pm\pi, \pm\pi)$. When non-Hermiticity is introduced, EPs emerge within the BZ (refer to Figs. \ref{fig1}(a) and (b)). These EPs are characterized by the coalescence of both eigenvalues and eigenvectors, and their loci are governed by the condition $\alpha(\mathbf{k}) = 0$ [\onlinecite{gkd2, gkd3}]. This implies that both the real and imaginary parts of $\alpha(\mathbf{k})$ must vanish simultaneously. Therefore, EPs are bound by the following condition

\begin{equation}
    \sum_{i}^{x,y}\sin^{2}(k_i) = \left(\frac{\xi}{\lambda}\right)^2.
\end{equation}
We highlight that the emergent EP is of the order of two, directly resulting from the square-root dispersion relation. The difference in the square-root eigenenergy of these EPs is accompanied by a branch cut that touches the Riemann sheet. We define the winding number of each EP around a path $\mathcal{S}^1$ parameterized by angle $\gamma \in [0,2\pi] $, given by [\onlinecite{wn}]
\begin{equation}
    W = \frac{-1}{2\pi}\Im\left[\int_{0}^{2\pi}\nabla_{\gamma}\log\det{[\mathcal{H}(\mathbf{k}(\gamma))]}.d\gamma\right].
\end{equation}

Consequently, these EPs are topologically characterized by quantized winding numbers with integer values as shown in Fig. \ref{fig1}(c). EPs are often accompanied by divergent fidelity susceptibility, a key indicator of quantum phase transitions or critical phenomena [\onlinecite{nfs1, nfs2, nfs3, nfs4, nfs5, nfs6, nfs7,PhysRevB.86.245424}]. To understand and quantify this divergence in our system, we analytically calculate the left and right eigenvectors $\ket{\psi^{L(R)}_{\pm}}$ of the Hamiltonian, which are given by

\begin{equation}
    \ket{\psi^{L(R)}_{\pm}} = C_{\pm}\begin{pmatrix}
        1 \\
        [g_\mathbf{k}^{L(R)}]_{\pm}
    \end{pmatrix},
\label{eqn8}
\end{equation}

where, the coefficients $[g_{\mathbf{k}}^{L(R)}]_{\pm}$ are given by
\begin{equation}
    [g_\mathbf{k}^L]_{\pm} = \frac{\Delta(\cos{k_x}-\cos{k_y}) - i\xi - E_{\pm}(\mathbf{k})^*}{\lambda(\sin{k_y} + i\sin{k_x})},
\end{equation}
and
\begin{equation}
    [g_\mathbf{k}^R]_{\pm} = \frac{\Delta(\cos{k_x}-\cos{k_y}) + i\xi - E_{\pm}(\mathbf{k})}{\lambda(\sin{k_y} + i\sin{k_x})},
\end{equation}
with $C_{\pm} = \frac{1}{1 + [g_{\mathbf{k}}^L]_{\pm}^*[g_{\mathbf{k}}^R]_{\pm}}$.

For a general perturbative Hamiltonian, $\mathcal{H}(s) = \mathcal{H}_0 + s\mathcal{V}$, where s is a tuning parameter, BFS $\chi_{\mathbf{k}}$ can be defined by [\onlinecite{nfs1,form}]
\begin{equation}
    \chi_{\mathbf{k}} = \frac{\bra{\Psi_{+}^{L}(\mathbf{k})}\mathcal{V}\ket{\Psi_{-}^{R}(\mathbf{k})}\bra{\Psi_{-}^{L}(\mathbf{k})}\mathcal{V}\ket{\Psi_{+}^{R}(\mathbf{k})}}{(E_{+}(\mathbf{k})-E_{-}(\mathbf{k}))^2},
    \label{eqn11}
\end{equation}

where $\ket{\Psi_{\pm}^{L(R)}}$ are the left(right) eigenfunctions of the unperturbed Hamiltonian $\mathcal{H}_0$ and $E_{\pm}$ corresponds to its eigenenergies. 

Considering the unperturbed model Hamiltonian $\mathcal{H}$ with perturbing potential $\mathcal{V} = i\delta\xi\sigma_z$ with $\lvert \delta\xi/\xi\rvert << 1$, the absolute value of BFS can be found by substituting the eigenvalues and eigenfunctions from Eqs. (\ref{eqn5}) and (\ref{eqn8}) into Eqn. (\ref{eqn11}). This yields the following

\begin{equation}
    \lvert\chi_{\mathbf{k}}\rvert = \left\lvert\frac{-\lvert C_+C_-\rvert^2(\delta\xi)^2}{4\alpha(\mathbf{k})}(1-\mathcal{F}_1(\mathbf{k}))(1-\mathcal{F}_2(\mathbf{k}))\right\rvert.
\end{equation}

where $\mathcal{F}_1(\mathbf{k}) = [g_{\mathbf{k}}^L]_{+}^*[g_{\mathbf{k}}^R]_{-}$ and $\mathcal{F}_2(\mathbf{k}) = [g_{\mathbf{k}}^L]_{-}^*[g_{\mathbf{k}}^R]_{+}$. Therefore, the BFS of the model Hamiltonian diverges (as $\alpha(\mathbf{k})\rightarrow 0$) when the system encounters one of the EPs in the BZ (see Fig. \ref{fig1}(d)).

\section{FEPs in periodically driven system}
We begin by examining a periodically quenched NH Hamiltonian operator given by
\begin{equation}
\mathcal{H}_{j} = \sum_{\mathbf{k}}\mathcal{G}_{j}(\mathbf{k}).\bm{\sigma}\ket{\mathbf{k}}\bra{\mathbf{k}},
\end{equation}
where $\mathcal{G}_{j}(\mathbf{k}) = d_j(\mathbf{k}) + i\xi_j(\mathbf{k})$ and $\mathbf{k}$ represents the quasi-momentum operator, spanning $[-\pi, \pi]$. The non-Hermiticity arises from the imaginary component, $\xi_j(\mathbf{k})$, which accounts for either on-site loss/gain or non-reciprocal hopping within the lattice.

Next, we consider a step driving protocol defined as
\begin{equation}
\mathcal{H}(t) = \begin{cases}
\mathcal{H}_{j=1},& t \in [mT, (m+1)T]\\
\mathcal{H}_{j=2}, & t \in [(m+1)T, (m+2)T].
\end{cases}
\end{equation}
Here, $\mathcal{H}_{j=1}$ and $\mathcal{H}_{j=2}$ denote the NH Hamiltonian in their respective driving windows, resulting in an effective drive period of $2T$ with $m \in \mathcal{Z}$. To simplify our analysis, we consider the one-period Floquet propagator operator, $ \hat{\mathcal{U}}_{2T} = \sum_{\mathbf{k}} \mathcal{U}_{2T} \ket{\mathbf{k}}\bra{\mathbf{k}}$, where 
\begin{equation}
\mathcal{U}_{2T} = \sum_{\mathbf{k}}e^{-i\mathcal{G}_{2}(\mathbf{k}).\bm{\sigma} T}e^{-i\mathcal{G}_{1}(\mathbf{k}).\bm{\sigma} T},
\label{eqn15}
\end{equation}
which satisfies the Floquet eigenvalue problem
\begin{equation}
\mathcal{U}_{2T}\ket{\Psi^{F,R}} = e^{-2i\epsilon (\mathbf{k})T}\ket{\Psi^{F,R}},
\end{equation}
where $\epsilon (\mathbf{k})$ and $\ket{\Psi^{F,R}}$ are the quasi-energies and right Floquet states of the effective Floquet Hamiltonian defind as $\mathcal{U}_{2T}= \text{exp}({- 2 i\mathcal{ H}_{F} T})$. Unlike their Hermitian counterparts, the Floquet operator is generally non-unitary, leading to complex quasi-energies $\epsilon (\mathbf{k})$. The effective Hamiltonian can therefore be expressed as
\begin{equation}
\mathcal{H}_{F} = \sum_{\mathbf{k}}\epsilon (\mathbf{k}) \hat{h}(\mathbf{k})\cdot \bm{\sigma} .
\label{eqn17}
\end{equation}

Using Eqs. (\ref{eqn15}) and (\ref{eqn17}), we obtain the quasi-energy written as
\begin{align}
& \cos{[2\epsilon (\mathbf{k})T}] = \cos{(\Xi_{1}(\mathbf{k})T)}\cos{(\Xi_{2}(\mathbf{k})T )}\nonumber \\
 & -(\sin{(\Xi_{1}(\mathbf{k})T)}\sin{(\Xi_{2}(\mathbf{k}) T})(\hat{\mathcal{G}}_{1}(\mathbf{k}).\hat{\mathcal{G}}_{2}(\mathbf{k}) ),
 \label{eqn19}
\end{align}
and the Floquet axis, $\hat h(\mathbf{k})= h(\mathbf{k})/\sin{[2\epsilon(\mathbf{k})T}]$ with $\sin{[2\epsilon(\mathbf{k})T}]= \lvert\lvert h(\mathbf{k}_0)\rvert\rvert$ and 
\begin{align}
& h(\mathbf{k}) = \cos{(\Xi_{2}(\mathbf{k})T )}\sin{(\Xi_{1}(\mathbf{k}) T)}\hat{\mathcal{G}}_{1}(\mathbf{k}) \nonumber \\
&+\cos{(\Xi_{1}(\mathbf{k})T)}\sin{(\Xi_{2}(\mathbf{k}) T)}\hat{\mathcal{G}}_{2}(\mathbf{k}) \nonumber \\
&+ (\sin{(\Xi_{1}(\mathbf{k})T)}\sin{(\Xi_{2}(\mathbf{k}) T})(\hat{\mathcal{G}}_{2}(\mathbf{k})\times\hat{\mathcal{G}}_{1}(\mathbf{k})).
\label{eqn18}
\end{align}

Here, $\Xi_{j}(\mathbf{k})= \sqrt{\sum_{i}\mathcal{G}_{j,i}^2(\mathbf{k})}$ and $\hat{\mathcal{G}}_{j}(\mathbf{k}) = \frac{\mathcal{G}_j(\mathbf{k})}{\Xi_{j}(\mathbf{k})}$ denote the modulus and unit direction of the complex vector $\mathcal{G}_{j}(\mathbf{k})$, respectively.  The Floquet operator $\mathcal{U}_{2T}$ is neither unitary nor Hermitian. This allows its complex eigenvalues, $\lambda(\mathcal{U}_{2T})$ and eigenvectors $\ket{\Psi_{F}^{R,L}}$ to be tuned to achieve EP degeneracies, a behavior distinctly different from their Hermitian counterparts.  Similarly, the corresponding Floquet Hamiltonian $\mathcal{H}_{F}$ is NH, meaning its fundamental quasienergies $\epsilon(\mathbf{k}) = \frac{i}{2T}\ln{\lambda(\mathcal{U}_{2T})}$ are neither purely real nor imaginary. In the unfolded-zone scheme, the full quasienergy spectrum is given by $\epsilon_n(\mathbf{k}) = \epsilon(\mathbf{k}) + n\frac{\pi}{2T}$. This framework is, therefore, broadly applicable to both classical and quantum systems with arbitrary dimensions, as it holds for any general Hamiltonian $\mathcal{H}(t)$. Furthermore, the system exhibits band degeneracy at quasi-energies $0$ and $\pi/2T$, corresponding to
\begin{equation}
\cos{[2\epsilon (\mathbf{k})T}] = \pm 1.
\label{eqn20}
\end{equation}

At an FEP, specifically at $\mathbf{k}=\mathbf{k}_0$, the Floquet eigenstates coalesce, rendering the Floquet propagator defective. Technically, this implies that the geometric multiplicity of the eigenvalues of the single-period propagator becomes less than their algebraic multiplicity at the FEP. To elucidate this phenomenon, we examine the behavior of the Floquet propagator at quasi-energies $\epsilon(\mathbf{k}_0)=0$ or $\pi/2T$ (or, more formally, where $\cos[2\epsilon(\mathbf{k}_0)T] = \pm 1$). At these points, the Floquet propagator is given by
\begin{equation}
\mathcal{U}_{2T}^{\epsilon (\mathbf{k}_0) = 0(\pi/2T)} = \pm 1 - ih(\mathbf{k}_0).\bm{\sigma}.
\label{eqn21}
\end{equation}
In the Hermitian limit, we note that $h(\mathbf{k}_0) = 0$. However, with the introduction of $\xi_j(\mathbf{k})$, $h(\mathbf{k}_0) \neq 0$, but its modulus reduces to
\begin{equation}
\lvert\lvert h(\mathbf{k}_0)\rvert\rvert = \sin{[2\epsilon (\mathbf{k}_0)T]} = 0.
\label{eqn22}
\end{equation}
Satisfying both Eqs. (\ref{eqn20}) and (\ref{eqn22}) allows band degeneracy to induce defectiveness, notably hosting FEPs in the Floquet operator at quasi-energies $0$ and $\pi/2T$. This leads to the coalescence of Floquet states at the FEP. To illustrate this, consider a simplistic case where $h(\mathbf{k}_0) = (1, \frac{i}{\sqrt{2}}, \frac{i}{\sqrt{2}})$, which satisfies Eq. (\ref{eqn22}). The non-unitary evolution operator then reduces to
\begin{equation}
\mathcal{U}_{2T}^{\epsilon (\mathbf{k}_0) = 0(\pi/2T)} = \pm 1 - i\begin{pmatrix}
\frac{i}{\sqrt{2}} & 1 + \frac{1}{\sqrt{2}} \\
1 - \frac{1}{\sqrt{2}} & - \frac{i}{\sqrt{2}}
\end{pmatrix}.
\end{equation}
This operator possesses a doubly-degenerate eigenvalue $\lambda(\mathcal{U}_{2T}) =  +1$ for $ \epsilon (\mathbf{k}_0) = 0$  and $\lambda(\mathcal{U}_{2T}) =  -1$ for $ \epsilon (\mathbf{k}_0) = \pi/2T$. The co-linear eigenvector for both cases has the form $v = [i(1+\sqrt{2}), 1]^T$ upto a normalization factor. This highlights a key characteristic of FEPs: the algebraic multiplicity of the eigenvalue exceeds its geometric multiplicity. Through a similarity transformation $P = [v, v^{'}]$ (where $v^{'}$ satisfies $(\mathcal{U}_{2T}-\mathcal{I})v^{'} = v$), the Floquet propagator can be reduced to its Jordan canonical form [\onlinecite{jb_EP}], written as following
\begin{equation}
 P^{-1}\mathcal{U}_{2T}P = \begin{pmatrix}
\pm 1 & 1 \\
0 & \pm 1
\end{pmatrix}.
\end{equation}
This explicitly demonstrates that the similarity transformation can turn it into a diagonal matrix and thus resulting in the defectiveness of the Floquet propagator and confirms the emergence of FEPs.

\begin{figure}
    \centering
    \includegraphics[width=1\linewidth]{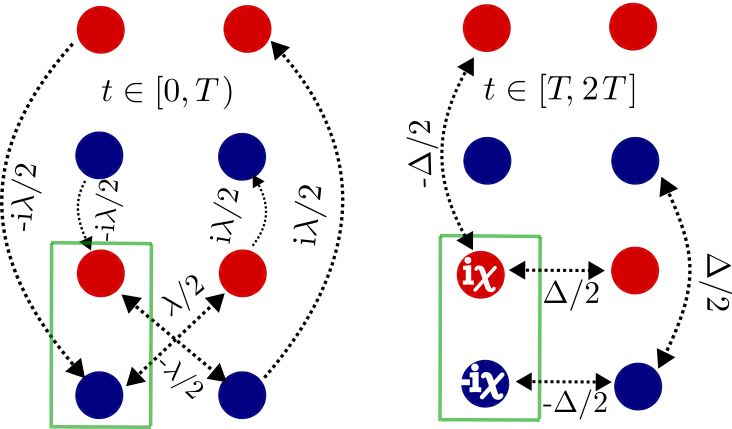}
    \caption{Realization of a periodically quenched Hamiltonian in a microcavity resonator. The diagram shows the system's evolution over two driving windows: an initial Hermitian window ($t\in[0,T)$) and a NH window ($t\in [T,2T)$). In the second window, non-Hermiticity is introduced through on-site gain (anti-clockwise modes, red dots) and loss (clockwise modes, blue dots). A green box highlights a specific lattice site, and the dotted arrows represent the coupling parameters $\Delta$ and $\lambda$.}
      \label{figl}
\end{figure}
\section{Periodically quenched driving}
To comprehensively understand the emergence of multiple FEPs and the higher-order skin effect, which arise from the intricate interplay between effective long-range hoppings induced by periodic driving and non-Hermiticity (imparting non-reciprocal character to these hoppings), we proceed with the two-dimensional periodically quenched Hamiltonian which is given by (see Fig. \ref{figl})

\begin{equation}
    \mathcal{H}_{j=1} = \lambda(\sin{k_x}\sigma_y - \sin{k_y}\sigma_x),
\end{equation}

and

\begin{equation}
    \mathcal{H}_{j=2} = (\Delta(\cos{k_x}-\cos{k_y}) + i\xi)\sigma_z,
\end{equation}

such that the effective Hamiltonian, in the static limit, precisely reduces to that of the two-band model encompassing multiple EPs in the BZ. (Refer to Eq. (\ref{eqn1}))

The system undergoes alternating periods of unitary and non-unitary evolution. Specifically, it evolves unitarily as $\mathcal{H}_{j=1}^{\dagger} = \mathcal{H}_{j=1}$ for $t \in [mT, (m+1)T]$  and non-unitarily as $(\mathcal{H}_{j=2}^{\dagger} \neq \mathcal{H}_{j=2})$ for $t \in [(m+1)T, (m+2)T]$  where $m$ is an integer. Although EPs are absent in either of the individual Hamiltonians, the behavior of driven system is different compared to that of the static case. We have seen it in the last section using a generic Hamiltonian containing FEPs. In the following, we use the quench drive and illustrate the generation of multiple FEPs. We begin by defining an effective NH Floquet Hamiltonian $\mathcal{H}_{F}(\mathbf{k}) = \frac{i}{2T}\ln{\mathcal{U}_{2T}(\mathbf{k})}$ (refer to Eqs. (\ref{eqn17})- (\ref{eqn19})). This is given by

\begin{equation}
   \mathcal{H}_{F}(\mathbf{k}) = \frac{\frac{1}{2T}\cos^{-1}{h_0(\mathbf{k})}}{\sin{[\cos^{-1}{h_0(\mathbf{k})}]}}\sum_{j}h_j(\mathbf{k})\sigma_j,
\end{equation}
where
\begin{equation}
    h_0(\mathbf{k}) = \cos{(\mathcal{G}T)}\cos{(\mathcal{G}_{2,z}T)},
\end{equation}

\begin{equation}
    h_x(\mathbf{k}) = \frac{\sin{(\mathcal{G}T)}}{\mathcal{G}}\left(\mathcal{G}_{1,x}\cos{(\mathcal{G}_{2,z}T)} - \mathcal{G}_{1,y}\sin{(\mathcal{G}_{2,z}T)}\right),
\end{equation}

\begin{equation}
    h_y(\mathbf{k}) = \frac{\sin{(\mathcal{G}T)}}{\mathcal{G}}\left(\mathcal{G}_{1,y}\cos{(\mathcal{G}_{2,z}T)} + \mathcal{G}_{1,x}\sin{(\mathcal{G}_{2,z}T)}\right),
\end{equation}

\begin{equation}
    h_z(\mathbf{k}) = \cos{(\mathcal{G}T)}\sin{(\mathcal{G}_{2,z}T)}.
\end{equation}

 Here $\mathcal{G}  = \lambda\sqrt{\sin^2{k_x} + \sin^2{k_y}}$, $\mathcal{G}_{1,x} = \lambda\sin{k_y}$, $\mathcal{G}_{1,y} = -\lambda\sin{k_x}$, and $\mathcal{G}_{2,z} = \Delta(\cos{k_x}-\cos{k_y}) + i\xi$. Therefore, the Hamiltonian hosts zero and $\frac{\pi}{2T}$ band degeneracy for $h_0(\mathbf{k}) = \pm 1$. (see Eq. (\ref{eqn20})) which leads us to the following relation
\begin{equation}
    (\cos{k_x}-\cos{k_y}) = \frac{\alpha\pi}{\Delta T},
\end{equation}
and
\begin{equation}
    \sqrt{\sin^{2}{k_x} + \sin^{2}{k_y}} = \frac{\beta\pi \pm \cos^{-1}(\frac{1}{\cosh{\xi T}})}{\lambda T}.
\end{equation}

\begin{figure}
    \centering
    \includegraphics[width=1\linewidth]{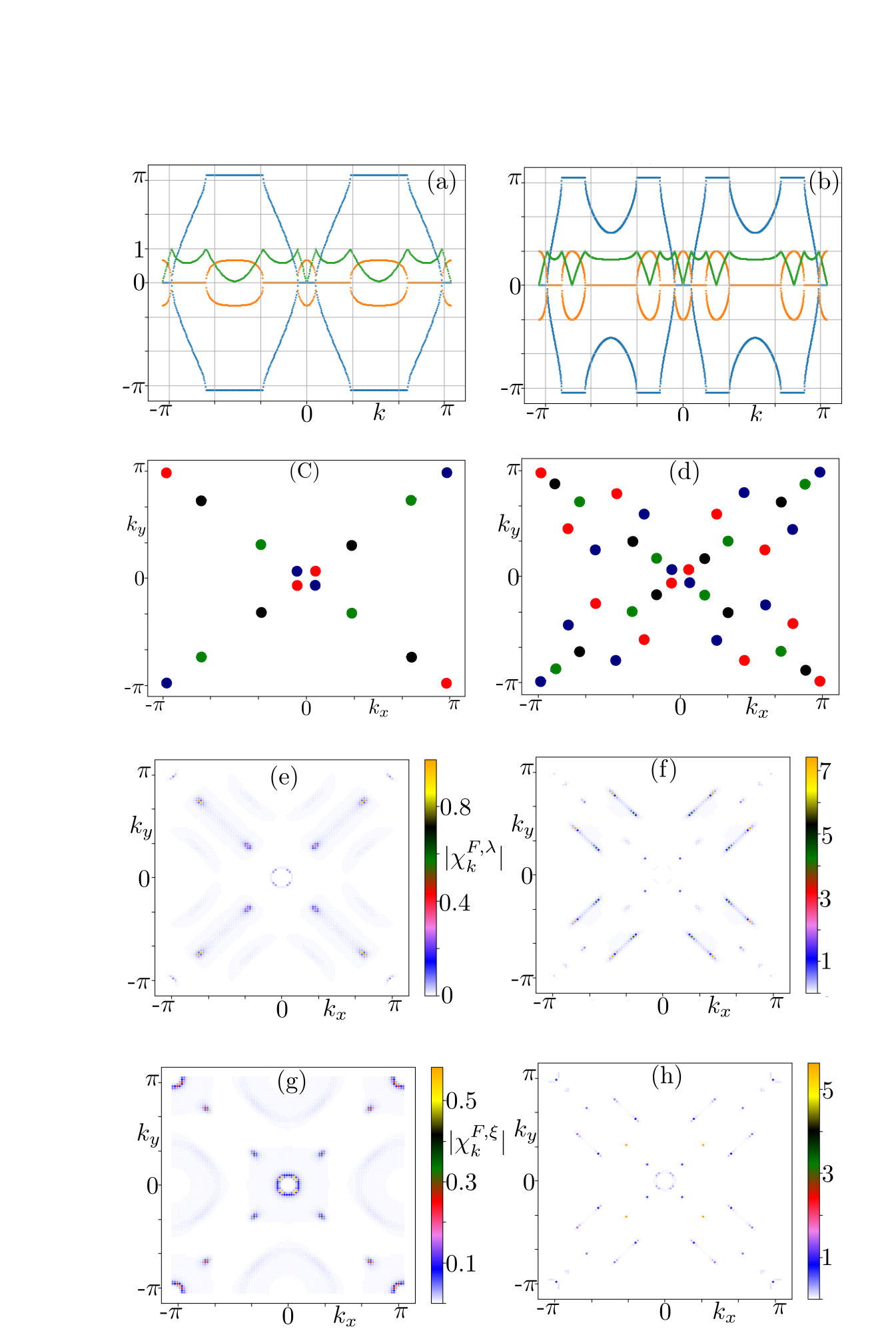}
    \caption{(a) and (b) present real (blue) and imaginary (orange) parts of the energy spectrum $2T\epsilon (k)$ with respect to the diagonal momentum ($k=k_x=k_y$). The overlap $\mathcal{O}( = \langle\Psi_{+}^{F,R}\lvert\Psi_{-}^{F,R}\rangle$) of right and left Floquet eigenstates of the effective Hamiltonian is superposed(green), confirming the emergence of FEPs. (c) and (d) show integer quantization of topological invariants $W^{0}$(for zero EPs) and $W^{\frac{\pi}{2T}}$(for $\pi$ EPs). For EPs at zero and $\pi$ quasi-energy, integer quantized values of +1 (-1) are denoted by red (blue) and green (black) respectively. (e) and (f) displays the absolute value of Floquet fidelity susceptibility $\chi_{k}^{F}$ which acquires non-zero value at the momenta hosting FEPs. The parameters are set as $\xi/\lambda = 0.3$ and $\Delta/\lambda = 1$ with $T=2.2$ for (a), (c) and (e) and $T=3.35$ for (b), (d) and (f).}
      \label{fig2}
\end{figure}

\begin{figure}
    \centering
    \includegraphics[width=1\linewidth]{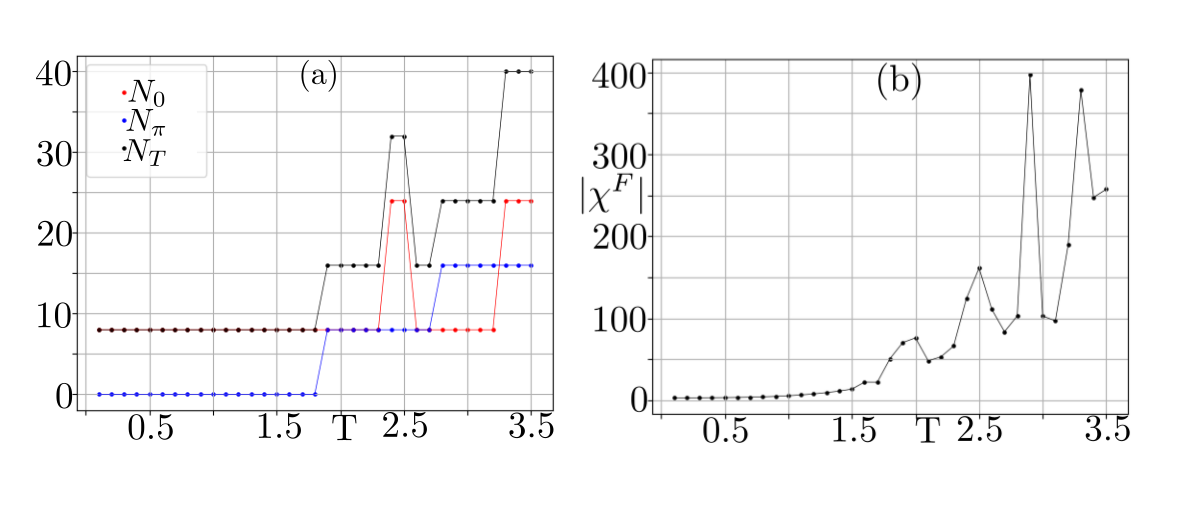}
    \caption{(a) shows the evolution of number of zero FEPs ($N_{0}$), $\pi$-FEPs ($N_{\pi}$), and total number of EPs ($N_T = N_0 + N_\pi$) with respect to the time period T showcasing the emergence of Type-II FEPs. (b) displays the divergence of summed over absolute FFS $\lvert\chi^{F}\rvert$ (written in Eq. (\ref{eqn37})) with respect to the half time period $T$ capturing the change of the number of FEPs. The parameters are set as $\xi/\lambda = 0.3$ and $\Delta/\lambda = 1$ with $\xi=0.3$. The values of the number of FEPs and summed over absolute Floquet fidelity susceptibility are evaluated with $\delta T = 0.1$ with the momentum grid $\delta k_{x(y)} = \pi/100$.}
      \label{fig4}
\end{figure}

  It is straightforward to see that the band degeneracy at zero ($\pi/2T$) quasi-energy corresponds to the case where $\alpha$ and $\beta$ are integers of the same (opposite) parity. Moreover, the band degeneracy at these quasi-energies exhibits multiple pairs of FEPs, which are characterized by coalescent Floquet eigenvectors. This shows the presence of periodically driven pair production and annihilation of FEPs (see Figs. \ref{fig2}(a) and (b)). Furthermore, these multiple FEPs can be classified into two distinct sub-classes. Firstly, Type-I FEPs manifest as a bifurcation of Floquet diabolic points, drawing a parallel to the emergence of EPs from Dirac points in static NH systems [\onlinecite{EP8,ft}]. Secondly, the model further reveals Type-II FEPs, which notably emerge as the time period of the driving field increases. Notably, these FEPs lack direct counterparts in static or periodically driven Hermitian diabolic point systems [\onlinecite{ft}]. The precise locations of these FEPs can be easily tuned by carefully modulating the system parameters. 
   We notice in Figs. \ref{fig4} that $\pi$-FEPs are being created at $\mathbf{k}_i = (\pm\frac{\pi}{2},\pm\frac{\pi}{2})$. Therefore, to simplify our demonstration for the $\pi$ Type-II FEPs, we restrict our analysis of to these momenta  and half-time period $T_i = [0,3.5]$. At these specific points, the Hermitian component of the effective Hamiltonian reduces to $\mathcal{H}_F^H(\mathbf{k}_i)=\frac{\pm\lambda}{2}(\sigma_y \pm \sigma_x)$ with the eigenvalues $E_{\pm}^{H}(\mathbf{k}_i) = \pm\frac{1}{\sqrt{2}}$. This explicitly demonstrates the absence of Floquet diabolic points in the Hermitian limit at these momenta. However, the FEPs manifest at these same momenta upon introducing non-Hermiticity which is bound by $\lvert\lvert h(\mathbf{k}_i)\rvert\rvert = 0$, see Eq. (\ref{eqn22}). This results into $\xi = \frac{1}{T}\tanh^{-1}\left[\pm \sin{\sqrt{2}\lambda T}\right]$ which underscores that Type-II FEPs inherently lack Hermitian diabolic point counterparts. Their precise temporal occurrences, at time period $T_i$, are analytically solved using Eq. (\ref{eqn20}) (setting 
$h(\mathbf{k}_i) = \pm 1$) which are governed by the following transcendental equation

  \begin{equation}
      T_i = \frac{1}{\xi}\mathcal{F}(\beta,T_i),
  \label{eqn31}
  \end{equation}
where $\mathcal{F}(\beta,T_i) = \cosh^{-1}(\frac{1}{\cos{\mathcal{A}(\beta,T_i)}})$ and $\mathcal{A}(\beta,T_i) = \sqrt{2}\lambda T_i - \beta \pi$. Numerical solution of the transcendental Eq. (\ref{eqn31}) allows for the determination of the precise temporal locations of zero and $\pi$ Type-II FEPs near $\mathbf{k}_i$. Notably, for zero-FEPs ($\beta=0$), no solutions are obtained at $\mathbf{k}_i$, confirming that their emergence is localized at different momenta. However, when considering $\pi$-FEPs ($\beta=1$), the computations provides distinct solutions: $T_i=1.85$ and $T_i=2.75$. These values unequivocally confirm the manifestation of $\pi$-FEPs at these specific half-time periods (Fig. \ref{fig4}(a)).

Next, we characterize these FEPs by calculating the winding number around a path $\mathcal{S}^1$ parameterized by angle $\gamma \in [0,2\pi] $ and given by
\begin{equation}
    W^{\epsilon} = \frac{-1}{2\pi }\Im\left[\int_{0}^{2\pi}\nabla_{\gamma}\log\det{[\mathcal{H}_{F}(\mathbf{k}(\gamma))- \epsilon]} d\gamma\right].
\end{equation}

We take the reference energy $\epsilon$ to be $0$ ($\pi/2T$) for the FEP located at quasi-energy zero ($\pi/2T$). These FEPs are topologically characterized by the winding number which can be related to the topological charge $\nu$ by the relation $W^{\epsilon} = -\nu/2$ [\onlinecite{tc1, tc2}]. The zero and $\pi$ EPs possess integer quantized winding numbers at different quasi-energies (see Fig. \ref{fig2}(c) and (d)). We also note that a NH counterpart of periodically driven QWZ model also showcases multiple FEPs in the BZ. (Refer Appendix A.)

To gain further insight, we define the BFFS $\chi_{\mathbf{k}}^{F,S}$ which is given by
\begin{equation}
    \chi_{\mathbf{k}}^{F,S} = \frac{\bra{\Psi_{+}^{F,L}(\mathbf{k})}\mathcal{V}^{F,S}\ket{\Psi_{-}^{F,R}(\mathbf{k})}\bra{\Psi_{-}^{F,L}(\mathbf{k})}\mathcal{V}^{F,S}\ket{\Psi_{+}^{F,R}(\mathbf{k})}}{(E_{+}(\mathbf{k})-E_{-}(\mathbf{k}))^2}.
\end{equation}
Here $\ket{\Psi_{\pm}^{F,L(R)}(\mathbf{k})}$ denotes the left (right) Floquet states  of the unperturbed  effective Hamiltonian $\mathcal{H}_F$ with $E_{\pm}(\mathbf{k})$ as its quasi-energies. Analogous to the static case, we perturbatively modify the parameter as $S \rightarrow S + dS$ where $\lvert dS/S\rvert << 1$. Consequently, the perturbing potential reduces to $\mathcal{V}^{F,S} = \mathcal{H}_{F}(S + dS)- \mathcal{H}_{F}(S)$, where $\mathcal{H}_{F}(S)$ represents the short-hand notation of the effective Hamiltonian $\mathcal{H}_{F}(\mathbf{k})$ with the parameter $S$. We note that the resulting BFFS, in general, is a tensor quantity. We focus on the case where $S = \lambda$ and $S=\xi$. The absolute value of these BFFS yield large non-zero values while encountering FEPs in the momentum space, thereby distinguishing the FEPs in the BZ (refer \ref{fig2}(e)-(h)).
To quantitatively characterize the emergence of FEPs as a function of increasing drive's time-period, we further define the momentum-summed absolute BFFS as
\begin{equation}
    \lvert\chi^{F}\rvert = \sum_{\mathbf{k}}\lvert\chi_{\mathbf{k}}^{F}\rvert.
    \label{eqn37}
\end{equation}

By varying the drive's half-time period $T$, the momentum-summed BFFS exhibits a sharp divergence precisely when FEPs emerge within the system. This observation conclusively confirms the appearance of multiple pairs of FEPs as a direct function of the drive's half-time period. (Refer to Fig. \ref{fig4} (b))

\begin{figure}
    \centering
    \includegraphics[width=1\linewidth]{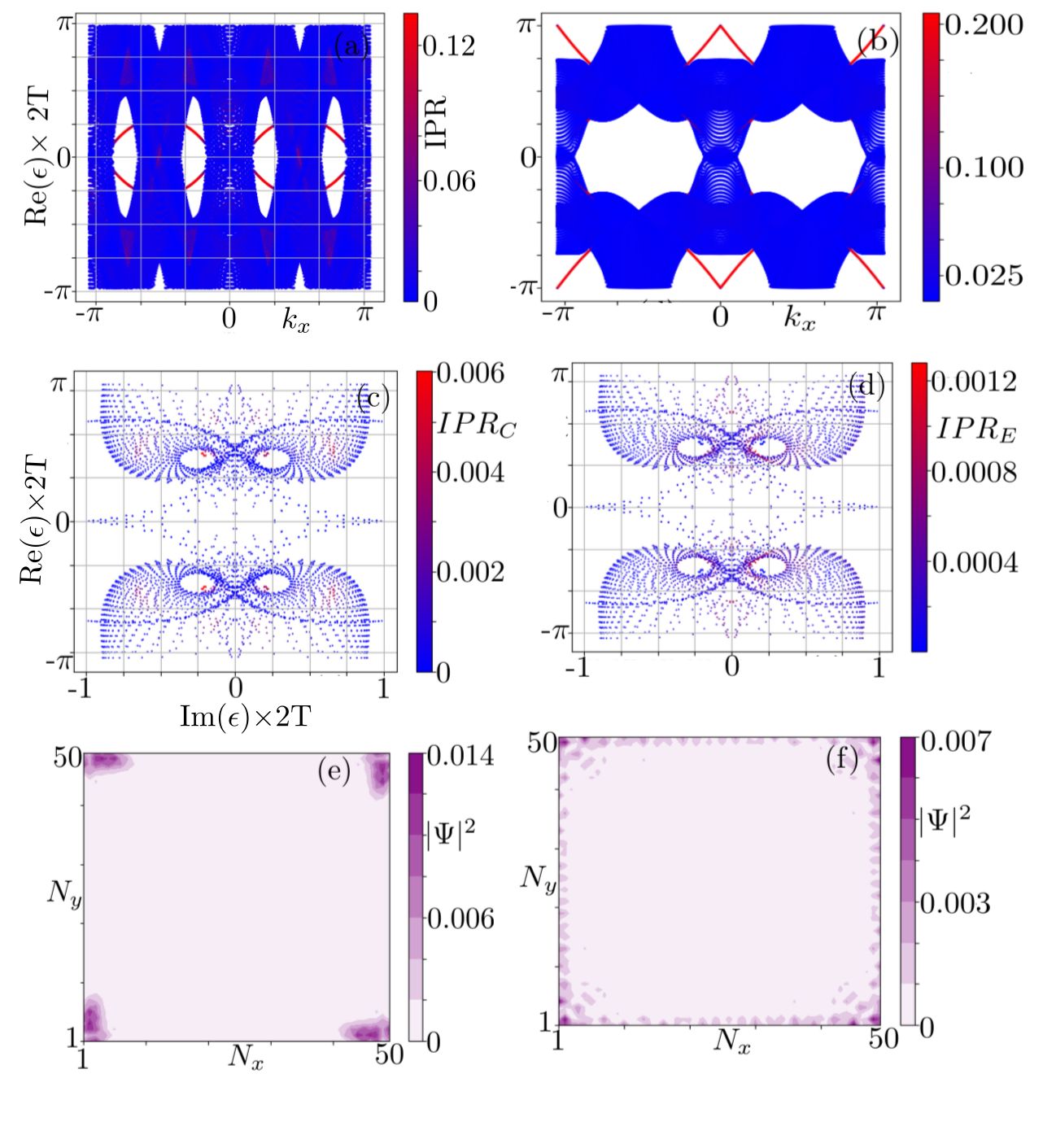}
    \caption{(a) and (b) depict the dispersion of the real energy $Re(\epsilon)$ of the effective Hamiltonian as a function of momentum $k_x$ under open boundary condition (OBC) along the $y$-direction, clearly showing the emergence of edge modes around zeo quasi-energy and a $\pi$-surface Dirac cone respectively. (c) and (d) reveal the emergence of corner and edge-localized skin modes under OBC along both axes. These modes are quantified by the corner-weighted ($\text{IPR}_C$) and edge-weighted ($\text{IPR}_E$) inverse participation ratios, respectively. (e) and (f) provide visual representations of the corner skin modes and edge skin modes, exhibiting localization at corners and edges, respectively. The parameters for (a)-(f) are set as $\xi/\lambda = 0.5$, $\Delta/\lambda = 1$ and $T=2$. $\xi$ is set to 1 and 0.5 for (a) and (b)-(f) respectively.}
    \label{fig3}
\end{figure}

The Hamiltonian now breaks the transposition-associated mirror symmetries ($\mathcal{M}_x$ and $\mathcal{M}_y$) while crucially retaining spatio-inversion symmetry ($\mathcal{P} = \sigma_z$). The absence of mirror symmetries further demands the presence of skin effect [\onlinecite{NHSE2}]. We note that in PBC, this Hamiltonian is characterized by four Floquet point gaps at quasi-energy $Re(\epsilon) = \pm\frac{\pi}{4T}$. However, when we consider open boundary conditions (OBC) along one of the axis, the edge modes around zero energy (see Fig \ref{fig3} (a))  and $\pi$ surface dirac cones (see Fig \ref{fig3}(b)) appear, confirming the emergence of first-order topology and its robustness against the NH perturbation.  Moreover, the model reveals the emergence of a higher-order skin effect when OBC are applied along both axes. This effect is characterized by the coexistence of corner and edge skin modes, which collectively define the higher-order skin effect. In order to quantitatively distinguish the emergence of these localized modes, we define the weighted Inverse Participation Ratio ($\text{IPR}_W$) [\onlinecite{NHSE4}] as

\begin{equation}
   \text{IPR}_{W} = \sum_{\tilde{r}}\sum_{i}e^{-\lvert r_i - \tilde{r}\rvert/\mathcal{R}}\lvert \Psi_{i}^{F}\rvert^{4}.
\end{equation}
Here, $\lvert \Psi_{i}^{F}\rvert^{4}$ quantifies the localization strength of the $i$-th Floquet eigenstate at all lattice sites. This term is multiplied by an exponential decaying factor that specifically captures the localization properties of the $W$ which can be either corner or edge modes. One also require that the decay length $\mathcal{R}$ to be significantly smaller than the system size ($\mathcal{R} << L$) to ensure accurate characterization of these localized states. We observe a Floquet higher-order skin effect, explicitly signified by the presence of corner modes not only in the point gaps of the quasi-energy spectrum but also within the bulk spectrum. Each of the point gaps in the quasi-energy spectrum supports two skin modes, which are four-fold degenerate and localized at the corners (see Figs. \ref{fig3}(c) and (e)). This effect is found to be robust as it is protected by spatio-inversion symmetry. In addition to the corner skin modes, the model also hosts edge-localized skin modes (see Figs. \ref{fig3}(d) and (f)). This coexistence of multiple skin modes signifies a transition to a state exhibiting a higher-order skin effect that is embodied by multiple EPs within the BZ.

\section{conclusion}
In this work, we have explored a two-dimensional, two-band NH system subjected to a periodically quenched driving protocol, revealing the fascinating interplay between non-Hermiticity and long-range hopping. This interplay, introduced by periodic driving, leads to the emergence of both FEPs and a higher-order NH skin effect. Our analysis shows that in the static limit, the model features eight EPs in the BZ, each characterized by an integer-quantized winding number, confirming their topological protection. A key finding is that the biorthogonal fidelity susceptibility exhibits sharp divergences at these momentum points, a defining characteristic of EPs in NH systems. We also highlight that such a system is experimentally realizable using microcavity resonators, specifically by introducing on-site loss and gain in the clockwise and anticlockwise modes, respectively.

Under the periodic driving protocol, the complex band structure of the system hosts multiple FEPs. We demonstrated that periodic driving facilitates the pair-production and annihilation of these EPs within the BZ. Notably, FEPs at zero and $\pi$ quasi-energies are characterized by their integer quantized winding numbers. While individual momentum points with FEPs show non-zero peaks in the biorthogonal Floquet fidelity susceptibility, the momentum-summed absolute value of this susceptibility exhibits a sharp divergence when the number of FEPs transitions. This clearly confirms the generation of multiple FEPs as the time period of the drive varies.
Furthermore, we have identified a higher-order NH skin effect in our model. The complex eigenspectrum of  effective Hamiltonian shows Floquet point gaps that specifically host corner-localized skin modes. These, along with edge-localized skin modes, collectively give rise to the higher-order skin effect.

Our findings provide a comprehensive understanding of the rich physics arising from NH Floquet systems, particularly the complex interplay of symmetries, topological features such as EPs, and higher-order skin effect. These insights pave the way for future research into designing and controlling NH topological phases in driven quantum systems.

\setcounter{equation}{0}

\begin{figure}
    \centering
    \includegraphics[width=1\linewidth]{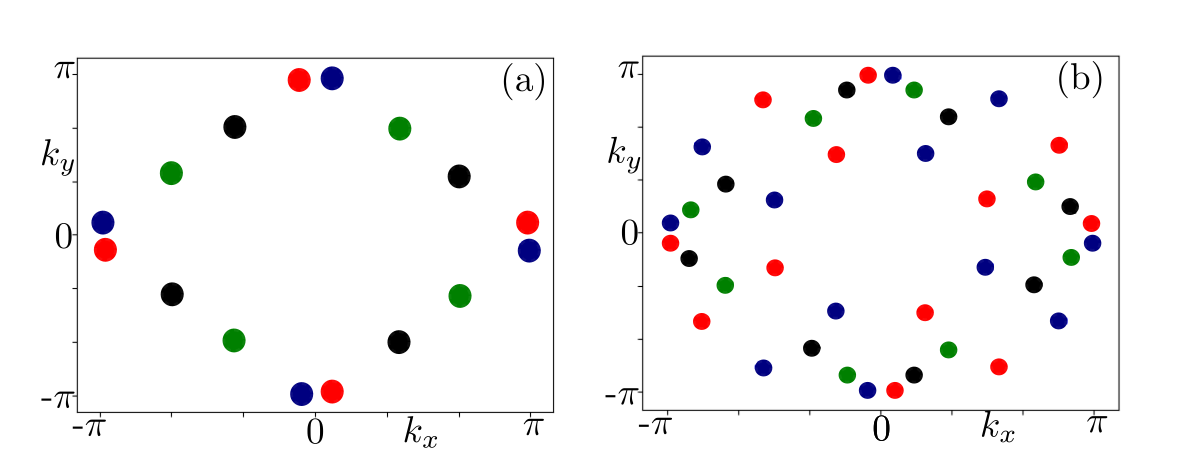}
    \caption{(a) and (b)  illustrate the integer quantization of topological invariants for the periodically quenched NH QWZ model. Specifically, $W^{0}$(for zero energy FEPs) and $W^{\frac{\pi}{2T}}$(for $\pi$ FEPs) are presented. For FEPs at zero and $\pi$ quasi-energy, integer quantized values of +1(-1) are denoted by red (blue) and green(black) markers respectively. The parameters are set as $\xi/\lambda = 0.3$ and $\Delta/\lambda = 1$. $T=2$ and $T=3.5$ for (a) and (b) respectively.}
      \label{figa}
\end{figure}
\section*{Appendix {A}}
\label{appA}
\setcounter{equation}{0}
For completeness, we also present a periodically quenched NH QWZ model showcasing multiple FEPs which is given by,
\begin{equation}
    \mathcal{H}_{j=1} = \lambda(\sin{k_x}\sigma_x + \sin{k_y}\sigma_y),
\end{equation}

and,

\begin{equation}
    \mathcal{H}_{j=2} = (\Delta(\cos{k_x}+\cos{k_y}) + i\xi)\sigma_z.
\end{equation}

The static limit of the model features band degeneracies that manifest as four pairs of EPs near momenta $(0,\pm\pi)$ and $(\pm\pi,0)$ in the BZ, each possessing an integer quantized winding number. In contrast, under periodic driving, the model gives rise to multiple FEPs. These FEPs can be analytically determined, following a similar methodology to that detailed in the main text. The zero and  $\pi$ FEPs are topologically characterized by integer quantized winding numbers. Consistent with earlier periodically quenched models, the momenta hosting these FEPs are characterized by a distinct non-zero peak in the absolute value of the BFFS. Moreover, the integrated absolute BFFS, summed over momentum, displays a sharp divergence as a function of the time period, indicative of Type-II FEP emergence.

\end{document}